\documentclass[aps,prl,superscriptaddress,showpacs,floatfix,twocolumn,graphicx]{revtex4-1}

\usepackage{amsmath}
\usepackage[dvipdfm]{graphicx}
\usepackage[colorlinks=true,linkcolor=blue,plainpages=false]{hyperref}
\usepackage{srcltx}
\usepackage{amsthm}

\def\v2{\mbox{$v_2$}}

\begin{document}

\title{\bf Intercepts of the momentum correlation functions in
$\mu$-Bose gas model and their asymptotics}

\author{
A.M.~Gavrilik$^1$\footnote{E-mail: omgavr@bitp.kiev.ua},
A.P.~Rebesh}

\affiliation{ Bogolyubov Institute for Theoretical Physics, 14b
Metrologichna Str.,
Kiev 03680, Ukraine \\
}
\date{\today}

\begin{abstract}
The so-called $\mu$-deformed oscillator (or $\mu$-oscillator)
introduced by A.\,Jannussis, though possesses rather exotic
properties with respect to other better known deformed oscillator
models, also has good potential for diverse physical applications.
     In this paper, the corresponding $\mu$-Bose gas model based on
$\mu$-oscillators is developed. Within this model, the intercepts
$\lambda^{(2)}(K)$ and $\lambda^{(3)}(K)$ of two- and three-particle
momentum correlation functions are calculated with the goal of
possible application for modeling the non-Bose type behavior of the
intercepts of two- and three-pion correlations, observed in the
experiments on relativistic heavy ion collisions. In derivation of
intercepts, a fixed order of approximation in the deformation
parameter $\mu$ is assumed. For the asymptotics of the intercepts
$\lambda^{(2)}(K)$ and $\lambda^{(3)}(K)$, we derive exact
analytical formulas.
 The results for $\lambda^{(2)}(K)$ are compared with experimental data, and
with earlier known results drawn using other deformed Bose gas
models.
\end{abstract}

  %% 02.20.Uw, 05.30.-d,


\maketitle
\section{\label{sec:1}1. Introduction}
For the last two decades, diverse models of deformed oscillators
have received much attention due to both their unusual properties,
as compared with the standard quantum harmonic oscillator, and their
great potential for being applied in description of miscellaneous
physical systems showing essential nonlinearities. The diversity of
applications ranges from, say, quantum optics and Landau problem to
high energy quantum particle phenomenology, modern quantum field
theory and D-brane (string) theory
\cite{manko95,Alvarez,Crnugelj,Geloun,Algin1,AG,Ribeiro,gavrS,Ko}.
Among the best known, extensively studied deformed oscillators we
encounter such as the $q$-deformed Arik-Coon (AC) one \cite{AC} and
Biedenharn-Macfarlane (BM) one \cite{B,M}, as well as the
two-parameter $p,\!q$-deformed oscillator \cite{chakjag}.
 Besides, there exists the $q$-deformed Tamm-Dancoff (TD)
 oscillator \cite{OKK,CSJ}, also explored though to much lesser extent \cite{GR1}.
 Unlike all these, there is very modest knowledge concerning
 the so-called $\mu$-deformed oscillator.
 Introduced in \cite{Jan} more than a decade and half
ago, the $\mu$-oscillator essentially differs from the models
already mentioned and exhibits
 rather unusual properties \cite{GKR}.

 An important direction of applying deformed oscillators
(deformed bosons) is the elaboration on their base of the respective
deformed versions of Bose gas model, like the $q$- or $p,q$-Bose gas
models in \cite{AGI1,G-NP,AGI2},\cite{AG,gavrS}. The $q$-deformed
extensions of Bose gas are used in modern approaches to $^4$He
theory, see e.g. \cite{Rego}. On the other hand, $q$-Bose and
$p,q$-Bose gas models provide an efficient ground for successful
description of the observed, in the experiments on heavy ion
collisions, unusual behavior of two-pion and three-pion momentum
correlations.
 As shown in \cite{AGI1,AGI2,ZP,AG,gavrS} the intercept
$\lambda^{(2)}(K)$, see formula (10) below, of the two-particle
momentum correlation function derived in the deformed Bose gas
models based on the AC- or BM-type $q$-deformed oscillators, as well
as on the $p,\!q$-oscillators, lead to a successful modeling of the
nontrivial, non-Bose type (such that basically differ from bosonic
case) properties of pion correlation intercepts in the RHIC and CERN
SPS experiments on relativistic collisions of nuclei.

Inspired by those results, in this paper we undertake an analogous
study basing on $\mu$-oscillators (\cite{Jan}, \cite{GKR}). Central
place in our treatment is given to calculation, within the
$\mu$-Bose gas model involving $\mu$-bosons ($\mu$-oscillators), of
the two- and three-particle correlation function intercepts
$\lambda^{(2)}(K)$ and $\lambda^{(3)}(K)$.
 For these we establish asymptotically constant type of behavior
({\it i.e.}, asymptotical approaching to a constant given solely by
$\mu$). In addition, we make some comparison of the obtained results
for intercepts with the existing data.

The paper is organized as follows. In Section 2 main facts about the
$\mu$-oscillator are given. For our goals we use the concept of the
{\it deformation structure function}
\cite{Meljanac,Bonatsos,Manko97} denoted as $\varphi(N)$. Here we
also sketch some unusual properties of the $\mu$-oscillator such as
the possibility of accidental degeneracy of energy levels and the
non-Fibonacci nature of its energy spectrum.
  Section 3 is devoted to calculation of the intercepts of
two- and three-particle momentum correlation functions.
 Namely, within the $\mu$-Bose gas model based on
$\mu$-oscillators we calculate these intercepts, for fixed order of
approximation.
 Derivation of their large momentum asymptotics which
is one of our main results is given in Section 4.
 Next 5th Section is devoted to a comparison with some experimental
data, and with analogous results obtained earlier within other ($q$-
or $p,q$-) deformed Bose gas models. The paper ends with concluding
remarks. Appendix A contains the data for the mean values or thermal
averages of diverse orders (for powers of the number operator), as
well as some interesting properties of the coefficients appearing in
the mean values.

\vspace{-4mm}
\section{\label{sec:2}2. Setup of the $\mu$-oscillator}
In the theory of deformed oscillators it is both convenient and
efficient to use the concept of {\it deformation structure
 function}, either in the $\varphi(N)$ version \cite{Meljanac,Bonatsos}
 or in the $f(N)$ version \cite{Manko97}.
 In our present treatment we prefer to use the former variant
$\varphi(N)$. Each specialized structure function determines the
particular deformed oscillator model through its corresponding
oscillator algebra, a unital algebra given in terms of the elements
$a$, $a^{\dagger}$, $N$.
 These generating elements of the algebra of a particular model
 of linear (usual) or nonlinear (deformed) quantum oscillator
 obey the defining relations
\begin{equation}
[a,a^{\dagger}]=F(N), \label{1}
\end{equation}
\begin{equation}
[N,a^{\dagger}]=a^{\dagger}, \hspace{15mm} [N,a]=-a.  \label{2}
\end{equation}
From (\ref{2}) it follows that
$[N,aa^{\dagger}]\!=\![N,a^{\dagger}a]\!=\!0$. The relation
(\ref{1}) involves $F(N)$, function of the number operator $N$. For
the usual quantum oscillator, $a^{\dagger}a\!=\!N$,
$aa^{\dagger}\!=\!N\!+\!1$ and $F(N)\!=\!1$. For general deformed
oscillator, like in \cite{Meljanac,Bonatsos}, we introduce the
deformation structure function $\varphi(N)$ according to the
relations
\begin{equation}
a^{\dagger}a=\varphi(N), \hspace{8mm} aa^{\dagger}=\varphi(N+1).
\label{3}
\end{equation}
Due to (\ref{1}) and (\ref{3}), the basic commutation relation takes
the form
\begin{equation}
aa^{\dagger}-a^{\dagger}a=F(N)=\varphi(N+1)-\varphi(N). \label{4}
\end{equation}
That is, the defining commutation relation is unambiguously given by
structure function. The Hamiltonian $H$ will be taken in the
standard form $H=\frac{1}{2}(aa^{\dagger}+a^{\dagger}a)$.
 Here and below we put $\hbar\omega=1$.
 In terms of structure function the Hamiltonian is
\begin{equation}
H=\frac{1}{2}\Bigl(\varphi(N+1)+\varphi(N)\Bigr). \label{5}
\end{equation}
We use the $q$-analog of Fock space. Within it, from the ground
state obeying
\[ a|0\rangle=0, \hspace{5mm} N|0\rangle=0,
\hspace{5mm} \langle0|0\rangle=1, \hspace{5mm}\]
 all the $n$-particle excited states are generated such that
\[
N|n\rangle=n|n\rangle, \hspace{5mm} \varphi(N)|n\rangle\ =
\varphi(n)|n\rangle.
\]
The $n$-particle {\it normalized} state is given by the formula
\[|n\rangle\!=\!\frac{(a^{\dagger})^n}{\sqrt{\varphi(n)!}}|0\rangle,
\hspace{3mm} \varphi(n)!\!=\!\varphi(n)\varphi(n-1)...\varphi(1),
\hspace{3mm} \varphi(0)!\!=\!1.\] The creation and annihilation
operators $a^{\dagger}$, $a$ when acting on the $n$-particle state
give
\[a^{\dagger}|n\rangle=\sqrt{\varphi(n+1)}|n+1\rangle,
\hspace{5mm} a|n\rangle=\sqrt{\varphi(n)}|n-1\rangle .\]

In what follows we focus on the so called $\mu$-deformed oscillator
\cite{Jan} for which the generating elements $a$, $a^{\dagger}$ and
$N$ of the corresponding algebra obey (\ref{2}) and (\ref{4}), and
the defining structure function is
\begin{equation}
\varphi(N)\equiv\varphi_{\mu}(N)=\frac{N}{1+\mu N} \label{6}
\end{equation}
where $\mu$ is the deformation parameter, $\mu\geq 0$. With this,
the basic relation and the energy spectrum for the Hamiltonian
(\ref{5}) of $\mu$-oscillator take the form
\begin{equation}
aa^{\dagger}\!-\!a^{\dagger}a\!=\!\varphi_{\mu}(N\!+\!1)\!-\!\varphi_{\mu}(N)\!=\!
\frac{N\!+\!1}{1\!+\!\mu(N\!\!+\!\!1)}\!-\!\frac{N}{1\!+\!\mu N}, \label{7}
\end{equation}
\begin{equation}
E_n=\frac{1}{2}\biggl(\frac{n+1}{1+\mu(n+1)}+\frac{n}{1+\mu
n}\biggr). \label{8}
\end{equation}
Setting $\mu=0$ recovers the algebra and the energies of the usual
harmonic oscillator.

Note that the linear structure function $\varphi(N)\!=\!N$, which
determines the standard quantum oscillator, stems as particular case
in the limit of deformation parameter(s), such as $q\rightarrow1$
for $q$-oscillators or $\mu\rightarrow0$ in the present case of
$\mu$-oscillator, from each model of deformed oscillators (a kind of
consistency).

\begin{center}
       {\it Unusual properties of $\mu$-oscillator}
\end{center}
\hspace{3mm}
 The ground state energy and the large $n$ asymptotics for the
energy of $\mu$-oscillator are respectively $E_0=\frac{1}{2(1+\mu)}$
and $E_\infty=\frac{1}{\mu}$. Both of these values basically differ
from the analogous value and limit for the usual oscillator,
$E_0\!=\!\frac{1}{2}$ and
$E_n\!\stackrel{n\to\infty}{\longrightarrow}\!\infty$. Moreover, the
ground state $E_0$ of the $\mu$-oscillator differs also from that of
the Fibonacci oscillators such as $p,\!q$-deformed family and its
one-parameter $q$-deformed subfamilies (AC, BM, TD), for which
zeroth level energy is $E_0\!=\!\frac{1}{2}$ exactly as for the
standard oscillator.

The just mentioned unusual feature of the $\mu$-oscillator is
tightly connected with the next its peculiarity: the
$\mu$-oscillator unlike the $p,\!q$-deformed one does not belong to
the class of Fibonacci oscillators. Namely, as shown in detail in
the recent paper \cite{GKR} the $\mu$-oscillator is typical
representative of the wide class of so-called quasi-Fibonacci
oscillators.

It is important that the $\mu$-oscillator admits nontrivial
extensions, in particular through combining it with other ($q$- or
$p,\!q$-) deformed oscillators. Any such extension will inherit the
peculiarities of those simpler deformed oscillators from which it is
composed, as shown in \cite{GKR}. As can be demonstrated for
positive $\mu$ such that $\mu<1$, the $\mu$-oscillator admits
occurrence of accidental energy level degeneracy if it is extended
by at least one deformation parameter. However, unlike the case of
Tamm-Dancoff $q$-deformed oscillator \cite{GR1}, here not all types
of pairwise energy level degeneracies are possible.

In other words, direct two- or three-parameter extensions of
$\mu$-oscillator can exhibit the phenomena of accidental degeneracy
(arising in the absence of any obvious underlying symmetry) of
energy levels. In more detail this property will be studied
elsewhere (see however \cite{GR3,GR1,GR4,GR2} where the ability to
display accidental energy levels degeneracy has been demonstrated
for the $p,\!q$-deformed and some $q$-deformed oscillators).

\vspace{-5mm}
\section{\label{sec:3}
3. The use of $\mu$-oscillators in $\mu$-Bose gas
model}\vspace{-3mm}
 Any real system of particles is far from being ideal.
 Therefore, a necessity arises to make complete account of
 inter-particle interactions.
Besides, the nonzero proper volume or substructure of particles may
also be of importance. For all these issues it is known that the
system admits effective description by means of deformed oscillators
\cite{Scarfone,Avancini,Perkins}. Due to that, the system can be
treated as an ideal one, but consisting of unconventional
appropriately deformed "bosons". As result, account of any of the
indicated complications or reasons turns out to be effectively
incorporated just in the particular deformation, with its specific
deformation parameter(s).

Such an effective picture which uses the $q$-deformed oscillators
has been applied to develop the related $q$-Bose gas model, with
subsequent successful application to superfluid $^4$He \cite{Rego}
or to an effective description \cite{AGI1,G-NP,AGI2,ZP} of the
observed \underline{non-Bose} type behavior of the intercept
$\lambda(m,\bf{K})$ of two-particle momentum correlation function of
pions produced and registered in the experiments on heavy-ion
collisions. "Non-Bose" means that the existing data for the
intercept clearly demonstrate the trend and the values that
principally differ (see {\it e.g.} \cite{Abelev,Humanic,AGP}) from
the would-be constant intercept equal to 1 at any momentum, for the
pions viewed as pure bosons.
  More advanced picture which involves the two-parameter $p,\!q$-deformed
oscillators and treats \underline{$n$-particle} correlation function
intercepts within the $p,\!q$-Bose gas model, along with analogous
application, has been developed in \cite{AG,gavrS}. Below we exploit
similar ideas for the case of $\mu$-Bose gas model.

\vspace{-5mm}
\subsection{\label{subsec:3.1}3.1. Intercept of two-particle
correlation function} \vspace{-3mm}
 In the experiments on heavy
ion-collisions, the colliding (primary) particles generate, as the
products of collisions, a miriads of secondary particles, {\it e.g.}
pions, which are then detected. As known \cite{LP,Heinz,Cs},
correlation functions carry important information concerning
geometry and dynamics of the emitting sources. The probability for a
joint observation of two particles with momenta $k_1$ and $k_2$ is
given by $P_2(k_1,k_2)$. The two-particle momentum correlation
function is defined, see {\it e.g.} \cite{LP,Heinz,Cs}, as
\begin{equation}
C^{(2)}(k_1,k_2)=\gamma\frac{P_2(k_1,k_2)}{P_1(k_1)P_1(k_2)}
\label{9}
\end{equation}
where $P_1(k_i)$, $i=1,2$, means pure single-particle probability
distribution for $i$-th particle and $\gamma$ is normalizing factor.
We put $\gamma=1$ in what follows. It is common to denote the
difference of 4-momenta of the pair by $Q=k_1-k_2$ and the mean
4-momentum by $K=(k_1+k_2)/2$. In the case of two identical (of mass
$m$) bosons and coinciding momenta, the $Q$ is vanishing. Then the
two-particle correlation function in variables $K$, $Q$, and the
reduced case $Q=0$ of it, is
\begin{equation}
C^{(2)}(Q,K)\stackrel{k_1=k_2}{\longrightarrow}C^{(2)}(Q\!=\!0,K)=
1+\lambda^{(2)}(m,\bf{K}) \label{10}
\end{equation}
where $\lambda^{(2)}(m,\bf{K})$ does appear.
 The entity $\lambda ^{(2)}(m,\bf{K})$ at an early stage was
 usually called the "coherence" or (non-) chaoticity parameter.
 On the other hand, since $C^{(2)}(Q,K)$ at $k_1=k_2$ attains its
maximal value, $\lambda ^{(2)}(K)$ is also termed the {\it
intercept} of the correlation function $C^{(2)}(Q,K)$ (and similarly
for $n$-th order correlation functions $C^{(n)}$). We will prefer to
use the latter term because, as stressed in \cite{gavrS}, even in
case of complete chaoticity of sources the intercept can attain the
 values differing from the would-be purely bosonic ones \cite{footnote}.
  The intercept can be expressed      \cite{LP}  in terms of particle
  (whose momentum is $K$) creation and annihilation operators as
\begin{equation}
\lambda^{(2)}(K)=\frac{\langle a^{\dagger}a^{\dagger}aa\rangle}{\langle
a^{\dagger}a\rangle^2}-1, \hspace{8mm} K=k_1=k_2. \label{11}
\end{equation}
A natural way to obtain after calculations the {\it nontrivial}
(non-constant) form of the intercept is to use some version of
deformed oscillator and to develop on this base the corresponding
deformed Bose gas model.

As mentioned, the explicit form of $\lambda^{(2)}(K)$ has been
earlier obtained in \cite{AGI1,G-NP,AGI2,GR4,AG} in the framework of
deformed versions of Bose gas model based on the popular AC-, BM-,
$p,\!q$-type deformed oscillator models. In the case of
$p,\!q$-bosons this result was extended \cite{AG,gavrS}  to the
$n$-particle correlation function intercepts $\lambda^{(n)}(K)$.

 Here we use the set of $\mu$-oscillators to develop the corresponding
$\mu$-Bose gas model. Within it, we derive the intercepts of the
two-particle and three-particle correlation functions. For that goal
we take (\ref{11}) as our starting point and in view of (\ref{3})
make account of the defining structure function of the chosen
   $\mu$-oscillator, namely
\begin{equation}
a^{\dagger}a=\varphi_{\mu}(N)=[N]_{\mu}\equiv\frac{N}{1+\mu N}\ .
\label{12}
\end{equation}
From eq. (\ref{2}), for a function $f(N)$ the property follows:
\begin{equation}
af(N)=f(N+1)a, \hspace{5mm} a^{\dagger}f(N)=f(N-1)a^{\dagger}.
\label{13}
\end{equation}
From (\ref{12}) we have one- and two-particle distributions
\[\hspace{-44mm}\langle a^{\dagger}a\rangle\!=
\!\langle [N]_{\mu}\rangle\!=\!\left\langle\! \frac{N}{1+\mu
N}\!\right\rangle,
\]\vspace{-5mm}
\begin{equation}
\hspace{-2mm}\langle a^{\dagger}a^{\dagger}aa\rangle\!=\!\langle
a^{\dagger}[N]_{\mu}a\rangle\!=\!\langle
a^{\dagger}a[N\!-\!1]_{\mu}\rangle\!=\!
\langle[N]_{\mu}[N\!-\!1]_{\mu}\rangle. \label{14}
\end{equation}
Let us point out the principal difference between deformed and
usual, non-deformed, approaches. Due to non-linear relation
(\ref{12}) of $a^{\dagger}a$ and the operator $N$ (for the usual
oscillator, we have $a^{\dagger}a=N$), one-particle distribution is
given by $\langle a^{\dagger}a\rangle$, and not by $\langle
N\rangle$.

Account of the both relations (\ref{14}) in eq. (\ref{11}) implies
\begin{equation}
\lambda^{(2)}_{\mu}(K)=\frac{\langle[N]_{\mu}[N-1]_{\mu}\rangle}{\langle
[N_{\mu}]\rangle^2}-1. \label{15}
\end{equation}
Assuming that $\mu$ is positive and sufficiently small we present
the ratio (\ref{12}) as \vspace{-1mm}
 \begin{equation} [N]_{\mu}\!=\! N-\mu
N^2+\mu^2N^3...=\!N\sum^{\infty}_{s=0}(-\mu N)^s \ . \label{16}
\end{equation}
After substitution of the expression (\ref{16}) in eq.~(\ref{15})
the formula for intercept of two-particle momentum correlation
function reads:
\begin{equation}  \label{17}
\lambda^{(2)}_{\mu}(K)=\!\frac{\langle
N(N\!-\!1)\sum_{k=0}^{\infty}(-\mu N)^k\sum_{r=0}^{\infty}(-\mu
(N\!-\!1))^r\rangle}{\langle\sum_{s=0}^{\infty}(-1)^s\mu^sN^{s+1}\rangle^2
}.
\end{equation}
Let us note that we cannot obtain single Taylor series for the ratio
$\lambda^{(2)}_{\mu}(K)$ in (\ref{15}), since the numerator and
denominator imply their own averages and, besides, we have no
general formula for necessary coefficients.

So, in calculating $\lambda^{(2)}_{\mu}(K)$ we deal with series and
hence use some approximation. Let us restrict ourselves with the
approximation in (\ref{16}) up to $\mu^5$.
 In this case the last formula (\ref{17}) can be cast to the form
\begin{equation}
\lambda^{(2)}_{\mu}(K)|_{\mu^5}=\frac{\sum^7_{r=1}\alpha_r\langle
N^r\rangle} {\bigl\langle
\sum^5_{s=0}(-1)^s\mu^sN^{s+1}\bigr\rangle^2}-1 \label{18}
\end{equation}
with the coefficients $\alpha_r$ as follows:

$\alpha_1=-1-\mu-\mu^2-\mu^3-\mu^4-\mu^5$,

\vspace{1mm} $\alpha_2=1+3\mu+4\mu^2+5\mu^3+6\mu^4+7\mu^5$,

\vspace{1mm} $\alpha_3=-2\mu-6\mu^2-10\mu^3-15\mu^4-21\mu^5$,

\vspace{1mm} $\alpha_4=3\mu^2+10\mu^3+20\mu^4+35\mu^5$,

\vspace{1mm} $\alpha_5=-4\mu^3-15\mu^4-35\mu^5$,

\vspace{1mm} $\alpha_6=5\mu^4+21\mu^5$,

\vspace{1mm} $\alpha_7=-6\mu^5$.

\noindent
 Next step of obtaining $\lambda^{(2)}_{\mu}(\overrightarrow{K})$
consists in finding mean values
of powers of $N$. Say, $\langle N\rangle$ yields (like in \cite{BB}):
\[
\langle N\rangle=\frac{Tr N e^{-\beta\sum_kH_k}}{Tr
e^{-\beta\sum_kH_k}}= \frac{\sum_n\langle n|N
e^{-\beta\sum_kH_k}|n\rangle}{\sum_n\langle
n|e^{-\beta\sum_kH_k}|n\rangle}=
\]
\vspace{-5mm}
\begin{equation}\hspace{-25mm}
=\frac{\sum_n ne^{-\beta\omega n}}{\sum_ne^{-\beta\omega n}}=
\frac{1}{e^{\beta\omega}-1} \label{19}\vspace{-1mm}
\end{equation}
where $\beta=\frac{1}{kT}$, $k=1$.
 Since in what follows we will use the developed model
for description of the intercepts of {\it pion} correlation
functions, we assume the particle's energy to be
 $\omega=
{(m^2+\bf{K}^2)^{1/2}}$.
 Similarly to (\ref{19}) we can derive the mean values $\langle N^k\rangle$,
$k\geq2$. Let us present those of them which are needed in
(\ref{18}): \vspace{-1mm}
\[\hspace{-7mm}
\langle N\rangle\!=\!\frac{1}{x}, \hspace{4mm}
 \langle N^2\rangle\!=\!\frac{1}{x}+\frac{2}{x^2}, \hspace{4mm}
\langle N^3\rangle\!=\!\frac{1}{x}+ \frac{6}{x^2}+\frac{6}{x^3},
\] \vspace{-4mm}
\[
\hspace{-44mm} \langle N^4\rangle\!=\!\frac{1}{x}+
\frac{14}{x^2}+\frac{36}{x^3}+\frac{24}{x^4},
\]\vspace{-4mm}
\[
\hspace{-30mm} \langle N^5\rangle\!=\!\frac{1}{x}+
\frac{30}{x^2}+\frac{150}{x^3}+ \frac{240}{x^4}+\frac{120}{x^5},
\]\vspace{-4mm}
\[\hspace{-16mm}
\langle N^6\rangle=\frac{1}{x}+\frac{62}{x^2}+
\frac{540}{x^3}+\frac{1560}{x^4}+\frac{1800}{x^5}+ \frac{720}{x^6},
\]\vspace{-4mm}
\[
\langle N^7\rangle=\frac{1}{x}+
\frac{126}{x^2}+\frac{1806}{x^3}+\frac{8400}{x^4}
+\frac{16800}{x^5}+\frac{15120}{x^6}+\frac{5040}{x^7}.
\]
The expressions for mean values of these and higher order powers
$N^m$ are of general form \vspace{-1mm}
\begin{equation}\label{34}
  \langle N^m\rangle\!=\!\sum_{r=1}^{m}B_r^{(m)}x^{-r},
    \hspace{14mm}  x\equiv e^{\beta \omega}-1. \vspace{-3mm}
 \end{equation}
 The coefficients $B_r^{(m)}$, both for $m\leq 7$ and for
$8\leq m\leq 14$, see Appendix A, show very interesting properties,
also mentioned in Appendix A: say, they are arranged in a remarkable
analog of the well-known Pascal's triangle.

 To summarize: the explicit expression for $\lambda
^{(2)}(m,\bf{K})$ in the $\mu^5$ approximation is given by eq.
(\ref{18}), with the above mean values $\langle N^k\rangle$. This
result covers the lower $\mu^3$ and $\mu^4$ orders as well. Using
data in Appendix A, similar result can be given for the order
$\mu^6$, and so on.

\vspace{-3mm}
\subsection{\label{subsec:3.2}3.2. Intercept of three-particle correlation function}
\vspace{-2mm} Like in the preceding Subsection, the intercept of
{\it three-particle} momentum correlation function is expressible in
terms of creation and annihilation operators: \vspace{-1mm}
\begin{equation}\vspace{-2mm}
\lambda^{(3)}_{\mu}(K)=\frac{\langle
a^{\dagger}a^{\dagger}a^{\dagger}aaa\rangle} {\langle
a^{\dagger}a\rangle^3}-1, \hspace{5mm} K=k_1=k_2=k_3. \label{20}
\vspace{-1mm}
\end{equation}
Using (\ref{13}) and (\ref{12})
the intercept $\lambda^{(3)}_{\mu}(K)$ takes the form
\begin{equation}
\lambda^{(3)}_{\mu}(K)=\frac{\langle[N]_{\mu}[N-1]_{\mu}[N-2]_{\mu}\rangle}
{\langle [N]_{\mu}\rangle^3}-1. \label{21} \vspace{-2mm}
\end{equation}
Account of (\ref{16}) in (\ref{21}) leads to general formula for
intercept of three-particle momentum correlation function:
\vspace{-6mm}
\begin{widetext}\vspace{-4mm}
\begin{equation}
\lambda^{(3)}_{\mu}(K)=\!\frac{\langle
N(N\!-\!1)(N\!-\!2)\sum_{m=0}^{\infty}(-\mu
N)^m\sum_{n=0}^{\infty}(-\mu (N\!-\!1))^n \sum_{r=0}^{\infty}(-\mu
(N\!-\!2))^r\rangle}{\langle\sum_{s=0}^{\infty}(-1)^s\mu^s
N^{s+1}\rangle^3}-1.  \label{22}
\end{equation}
\end{widetext}
Like before, we consider the order $\mu^5$ approximation and obtain
the expression for $\lambda^{(3)}_{\mu}(K)$ in the form
\vspace{-1mm}
\begin{equation}
\lambda^{(3)}_{\mu}(K)|_{\mu^5}=\frac{\sum^{8}_{k=1}\eta_k\langle
N^k\rangle} {\bigl\langle
\sum^{5}_{r=0}(-1)^r\mu^rN^{r+1}\bigr\rangle^3}-1 \label{24-}
\end{equation}
where the coefficients $\eta_k$ are the following:

\vspace{2mm} $\eta_1=2+6\mu+14\mu^2+30\mu^3+62\mu^4+126\mu^5$,

\vspace{2mm} $\eta_2=-3-15\mu-45\mu^2-115\mu^3-273\mu^4 -623\mu^5$,

\vspace{2mm} $\eta_3=+1+12\mu+55\mu^2+180\mu^3+511\mu^4+1344\mu^5$,

\vspace{2mm} $\eta_4=-3\mu-30\mu^2-145\mu^3-525\mu^4-1652\mu^5$,

\vspace{2mm} $\eta_5=6\mu^2+60\mu^3+315\mu^4+1260\mu^5$,

\vspace{2mm} $\eta_6=-10\mu^3-105\mu^4-602\mu^5$,

\vspace{2mm} $\eta_7=15\mu^4+168\mu^5$,

\vspace{2mm} $\eta_8=-21\mu^5$.

\vspace{-3mm}
\section{\label{sec:4}4. Asymptotics of intercepts of two- and
                         three-particle correlation functions}\vspace{-4mm}
 Here we study the asymptotical behavior at $\beta\omega\rightarrow\infty$
 ({\it i.e.} at $|\bf{K}|\to\infty$, constant $T$) of the intercepts
$\lambda_\mu^{(2)}$ and $\lambda_\mu^{(3)}$ of the two- and
three-particle correlation functions.

 Let us find the asymptotical expression for $\lambda^{(2)}_{\mu}$
within the order $\mu^5$ approximation, see (\ref{18}). With the
account of relevant averages from Appendix A both in
 the numerator and in the denominator of (\ref{18}), we derive the result
\begin{equation}  \label{23}
\lambda^{(2)}_{as.}(\mu)|_{\mu^5}=\frac{1-4\mu+11\mu^2-26\mu^3+
57\mu^4-120\mu^5}{1-2\mu+3\mu^2-4\mu^3+5\mu^4-6\mu^5}\ .
\end{equation}
This expression can be written in the form
\begin{equation}  \label{023}
\lambda^{(2)}_{as.}(\mu)|_{\mu^k}=
\frac{\sum^k_{s=0}(-1)^s(2^{s+2}-s-3)\mu^{s}}{\sum^k_{r=0}(-1)^r(r+1)\mu^r}
\end{equation}
where $k=5$. Obviously, the orders $\mu^3$ and $\mu^4$ are also
covered by the latter formula at $k=3$ and $k=4$, in agreement with
remark in the last paragraph of sect. 3.1 (about eq.(\ref{18}) and
the orders $\mu^3$, $\mu^4$).
 By calculation, we have verified that (\ref{023}) is also true at $k=6$,
and there is confidence in its validity for any $k<\infty$.
   On this base, after necessary summations we arrive at
  the {\it exact asymptotical formula}
\[\hspace{-11mm}
\lambda^{(2)}_{as.}(\mu)%|_{\mu^k}\!=
=\!\frac{\sum^{\infty}_{s=0}(-1)^s
(2^{s+2}\!-\!s\!-\!3)\mu^s}{\sum^{\infty}_{r=0}(-1)^r(r\!+\!1)\mu^r}\!=
\]\vspace{-5mm}
\begin{equation}
=\frac{1}{(1+\mu)^{2}(1+2\mu)}\Bigl(\frac{1}{(1+\mu)^2}\Bigr)^{-1} =
\frac{1}{1+2\mu}\ . \label{24}
\end{equation}

It should be noted that in the asymptotics the intercept does not
depend on the temperature, nor on the momentum and mass of particle.
It is the deformation parameter $\mu$ alone that determines the
intercept asymptotics.  So $\lambda^{(2)}_{as.}(\mu)$ is a number
$<1$ at any fixed $\mu>0$.

 Now consider the asymptotics of the {\it
three-particle} correlation function intercept.
 Applying the limit $\beta\omega\rightarrow \infty$ to eq. (\ref{24-}), with
the $\langle N^s\rangle$ taken from Appendix A, we obtain the
desired asymptotic formula in $\mu^5$ approximation:
\begin{equation}
\lambda^{(3)}_{as.}(\mu)|_{\mu^5}\!=\!\frac{6(1\!-\!6\mu+25\mu^2-90
\mu^3\!+\!301\mu^4\!-\!966\mu^5)}{1\!-\!3\mu\!+\!6\mu^2\!-\!10\mu^3\!+
\!15\mu^4\!-\!21\mu^5}\!-\!1.
\end{equation}
Using this expression (which also contains the order $\mu^3$,
$\mu^4$ results)
we find the {\it exact asymptotical} formula  % for  of $\lambda^{(3)}$
\[
\lambda^{(3)}_{as.}(\mu)%|_{\mu^5}\!
=\!\frac{6\sum^{\infty}_{s=0}(-1)^s
\Bigl(\frac12(3^{s+2}\!+\!1)\!-\!2^{s+2}\Bigr)\mu^s}
{\sum^{\infty}_{r=0}(-1)^r\frac12(r\!+\!1)(r\!+\!2)\mu^r}-1=
\]\vspace{-3mm}
\begin{equation}
=\frac{\frac{6}{(1+\mu)(1+2\mu)(1+3\mu)}}{\frac{1}{(1+\mu)^3}}-1=
\frac{5+7\mu}{(1+2\mu)(1+3\mu)}. \label{28}
\end{equation}
In fig.~1 we present the two- and three-particle correlation
function intercepts $\lambda^{(2)}_{\mu}$, $\lambda^{(3)}_{\mu}$ for
the order $\mu^5$ approximation, with their asymptotes shown
explicitly.
  \vspace{1mm}
 \begin{figure}[h]
\hspace{-75.6mm}
\includegraphics[angle=0,width=0.90\textwidth]{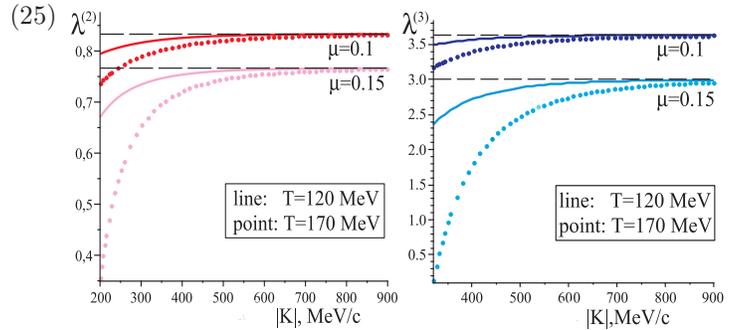}
\vspace{-15mm}
  \caption{\label{Fig:1}{{\em Left}:
  The intercept $\lambda^{(2)}$ to order $\mu^5$, with
asymptotes (dashed lines). Asymptote for $\mu$=0.1 curves is
$\lambda^{(2)}_{as.}=0.8331$ and for $\mu$=0.15 curves is
$\lambda^{(2)}_{as.}=0.7664$.
 {\em Right}: The intercept $\lambda^{(3)}$ to the order
 $\mu^5$, with asymptotes.
 Asymptote for $\mu$=0.1 curves is
$\lambda^{(3)}_{as.}=3.6365$ and for $\mu$=0.15 curves is
$\lambda^{(3)}_{as.}=2.9964$.}} \hspace{-7mm}
\end{figure}

\vspace{-6mm}
\section{\label{sec:5}5. Intercepts in $\mu$-Bose gas model compared with other models
and experimental data}\vspace{-3mm}
  It is of interest to compare the behavior of
the intercepts $\lambda^{(2)}$ and $\lambda^{(3)}$ obtained in the
$\mu$-Bose gas, see fig.~1, with main features of such
 intercepts found within $q$-Bose gas models
 \cite{gavrS,AGI1,G-NP,AGI2,GR4} and plotted in the corresponding figures
of those works.  As seen from the comparison, the picture is
qualitatively the same: with increasing mean momentum, in each model
we observe the occurrence of minimum, convexity downwards, then the
transition through inflection point to the convexity upwards, and
finally asymptotical tending to a constant value strictly lesser
than 1 if $q\neq 1$ or $\mu\neq 0$.

Now consider some experimental data, {\it e.g.}, taken from
\cite{Abelev}. It is nice that the experimental data for the
two-pion correlation intercepts (4 points corresponding to 4
momentum bins) respect the basic features mentioned in the previous
paragraph (the data show stable qualitative picture or "trend", for
different colliding nuclei, energies of collision and centralities).
 In our fig.~2 ({\it left}), for illustration, we make a comparison of
the curve $\lambda^{(2)}$ from (\ref{18}) with experimental points
\cite{Abelev} corresponding to the most central 62.4 GeV Au+Au
collision. In principal, we could get besides qualitative, also
better quantitative agreement with data within the considered
$\mu$-Bose gas model, although by taking lesser temperature and
slightly greater value of $\mu$. Note again that the fig.~2 ({\it
left}) serves for illustration only, and more detailed and realistic
comparative analysis will be given elsewhere.

What concerns the existing data for three-pion correlation
intercept, these are obviously insufficient. For that reason we
postpone confronting with data of our analytical results on
$\lambda^{(3)}(K)$ for future work as well.

Before ending this section, let us note that an important function
$r^{(3)}(K)$ was introduced in \cite{HeinzZ}. It is composed from
$\lambda^{(2)}(K)$ and $\lambda^{(3)}(K)$ according to the formula
\begin{equation}
r^{(3)}(K)=\frac12 \frac{\lambda^{(3)}(K) - 3
\lambda^{(2)}(K)}{(\lambda^{(2)}(K))^{3/2}}. \label{029}
\end{equation}
Due to special form of the ratio, the function (\ref{029}) provides
improved purity (on the experimental side). Indeed, many unwanted
distortions which especially for low $K$ may affect
$\lambda^{(2)}(K)$ and $\lambda^{(3)}(K)$ when considered
separately, in $r^{(3)}(K)$ are mutually canceled.

In case of $q$-Bose gas this function was studied in \cite{gavrS} by
one of us. From our analytical results (\ref{18}) and (\ref{24-})
derived in the $\mu$-Bose gas model, the formula for the function
$r_\mu^{(3)}(K)$ reads
\begin{equation}
r_\mu^{(3)}(K)=
 \frac{\frac{\sum^{8}_{l=1}\eta_l\langle
N^l\rangle} {\bigl\langle
\sum^{5}_{s=0}(-1)^s\mu^sN^{s+1}\bigr\rangle^3} - 3
\frac{\sum^{7}_{k=1}\alpha_k\langle N^k\rangle} {\bigl\langle
\sum^{5}_{r=0}(-1)^r\mu^rN^{r+1}\bigr\rangle^2} + 2}{ 2
\biggl(\frac{\sum^{7}_{k=1}\alpha_k\langle N^k\rangle} {\bigl\langle
\sum^{5}_{r=0}(-1)^r\mu^rN^{r+1}\bigr\rangle^2} -1 \biggr)^{3/2}}.
\label{29}
\end{equation}
Finally, using exact asymptotical formulas (\ref{24}) and (\ref{28})
we find the {\it exact expression} for the $r_\mu^{(3)}(K)$
asymptotics:
\begin{equation}
r_{as.}^{(3)}(\mu)=\frac{1-\mu}{1+3\mu}\sqrt{1+2\mu}. \label{31}
\end{equation}
Fig.~2, {\it right}, gives the plot of $r_\mu^{(3)}(K)$ along with
its asymptotics.
 As seen, small raise of the parameter $\mu$ results in basic
 change of the properties of $r_\mu^{(3)}(K)$: from positive to
 negative sign. But, as seen from (\ref{31}) its asymptotical
 value is always positive since $\mu<1$.
\begin{figure}[h]
  \includegraphics[angle=0,width=0.50\textwidth]{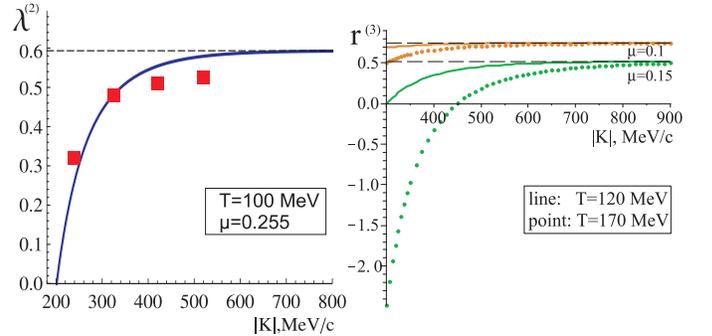}
  \caption{\label{Fig:1}{{\em Left}: The intercept
$\lambda^{(2)}_\mu(K)$ from (\ref{18}), to $\mu^5$, versus
experimental data \cite{Abelev}.  The parameter values are
$\mu=0.255$, $T=100$ MeV. \
 {\em Right}: The function $r_\mu^{(3)}(K)$ from (\ref{29}), to $\mu^5$,
 with its asymptotes (dashed horizontal lines).
  Asymptote for the $\mu$=0.1 curves is
$r^{(3)}_{as.}=0.7472$  and for the $\mu$=0.15 curves is
$r^{(3)}_{as.}=0.5194$.}}
\end{figure}

\vspace{-8mm}
\section{\label{sec:6}6. Conclusions}\vspace{-3mm}
 The $\mu$-deformed oscillator model is a peculiar one
possessing novel interesting features with regards to the other,
more popular and better known models of deformed oscillators. The
specifics of the $\mu$-oscillator is rooted in its defining
structure function whose non-polynomial, rational dependence on the
number operator $N$ causes nontrivial properties of this oscillator
especially those mentioned at the end of sect. 2.

The energy spectrum of $\mu$-oscillator does not form \cite{GKR} the
usual Fibonacci sequence. For that reason we treated the
$\mu$-oscillator as the one obeying a generalization of Fibonacci
relation and called it quasi-Fibonacci oscillator \cite{GKR}.
 We hope that the deformed oscillators possessing quasi-Fibonacci
property may be useful not only in the present context of
multiparticle correlations of pions (kaons) produced and registered
in the experiments at RHIC or LHC, but also, say, for description of
the properties of some quasi-periodic chains \cite{Negro-Feng}.

In the present paper, the $\mu$-oscillators are employed for
elaborating on their base the corresponding $\mu$-Bose gas model.
 The results  concerning one-, two-particle
distribution functions and the two-particle correlation function
intercept obtained in $\mu$-Bose gas model are derived by using
definite ($\mu^5$) order of approximation. From the obtained
intercept of two-particle correlation function depending on the mean
momentum $K$, the exact analytical formula for its asymptotics is
found. We observe the important fact that in the asymptotics, the
intercept does not depend on the particle mass, nor on the
temperature or momenta: the deformation parameter $\mu$ is the only
entity that survives and completely determines the
$\beta\omega\to\infty$ asymptotics of $\lambda^{(2)}(\mu)$.
Analogous set of results, {\it i.e.} the explicit formulas and the
exact asymptotical expression have been obtained for the intercept
$\lambda^{(3)}$ of three-particle correlation function.

Let us emphasize ones more that the obtained analytical formulas for
$\lambda_\mu^{(2)}(K)$, $\lambda_\mu^{(3)}(K)$ and $r_\mu^{(3)}(K)$
provide explicit and nontrivial dependence on the particles' mean
momentum.
 This allows to make a comparison with experimental
data on intercepts of pions, as shown in special case, see fig.~2,
{\it left}. Certainly, more detailed data are needed in order to
make choice in favor of a particular model and we believe the
situation will soon improve.

We hope that the above $\mu$-Bose gas model constructed by employing
$\mu$-oscillators will find interesting applications in different
branches of nonlinear quantum physics: in description of high-energy
quantum physics phenomena, in $^4$He theory, in quantum optics,
  {\it etc}.

\vspace{-3mm}
\section{Acknowledgments}\vspace{-2mm}
The authors are thankful to the referees for valuable remarks and
recommendations. This research was partially supported by the Grant
29.1/028 of the State Foundation of Fundamental Research of Ukraine
and by the Special Program of the Division of Physics and Astronomy
of the NAS of Ukraine.

\begin{widetext}

\vspace{-8mm}
\section{Appendix A.\ \
 Explicit expressions for the mean values $\langle N^m\rangle$
 }
 \vspace{-2mm}
 The expressions for mean values of the powers $N^m$
 are given as \ $\langle
 N^m\rangle\!=\!\sum_{r=1}^{m}B_r^{(m)}x^{-r},
    \hspace{3mm}
x\equiv e^{\beta \omega}-1$,
 with the coefficients $B_r^{(m)}$, \ $m\!=\!8,...,14$, placed in the
Table 1. For $m\!=\!1,...,7$ see the expressions preceding eq.
(\ref{34}).

\begin{center}
\bf{\tablename{ 1. The coefficients $B_r^{(m)}$ for mean values of
the powers $N^m$\ ($8 \le m\le 14$) }}\vspace{3mm}

{\begin{tabular}{|c|c|c|c|c|c|c|c|}%
 \hline
  ${\it r\setminus m}$ & {\it 8} & {\it 9} & {\it 10} & {\it 11} & {\it 12}
  & {\it 13} & {\it 14}\\
\hline
${\it 1}$ & $1$ & $1$ & $1$ & $1$ & $1$ & $1$ & $1$  \\
\hline ${\it 2}$ & $254$ & $510$ & $1022$ & $2046$ & $40946$ &
$8190$ & $16382$\\ \hline ${\it 3}$ & $5796$ & $18150$ & $55980$ &
$171006$ & $519156$
 & $1569750$ & $4733820$\\
\hline ${\it 4}$ & $40824$ & $186480$ & $818520$ & $3498000$ &
$14676024$ & $60780720$ & $249401880$ \\ \hline ${\it 5}$ & $126000$
& $834120$ & $5103000$ & $29607600$ & $165528000$ & $901020120$ &
$4809004200$ \\
\hline ${\it 6}$ & $191520$ & $1905120$ & $16435440$
& $129230640$ & $953029440$ & $6711344640$ & $45674188560$ \\
\hline ${\it 7}$ & $141120$ & $2328480$ & $29635200$ & $322494480$ &
$3162075840$ & $28805736960$ & $248619571200$ \\
\hline ${\it 8}$ & $40320$ & $1451520$ & $30240000$ & $479001600$ &
$6411968640$ & $76592355840$ & $843184742400$ \\
\hline ${\it 9}$ & $--$ & $362880$ & $16329600$ & $419126400$ &
$8083152000$ & $130456085760$ & $1863435974400$ \\
\hline ${\it 10}$ & $--$ & $--$ & $3628800$ & $199584000$ &
$6187104000$ & $142702560000$ & $2731586457600$ \\
\hline ${\it 11}$ & $--$ & $--$ & $--$ & $39916800$ & $2634508800$ &
$97037740800$ & $2637143308800$ \\
\hline ${\it 12}$ & $--$ & $--$ & $--$ & $--$ & $479001600$ &
$37362124800$ & $1612798387200$ \\
\hline ${\it 13}$ & $--$ & $--$ & $--$ & $--$ & $--$ & $6227020800$
& $566658892800$ \\
\hline ${\it 14}$ & $--$ & $--$ & $--$ & $--$ & $--$ & $--$ &
$87178291200$ \\
\hline
\end{tabular}}
\end{center}
\vspace{1mm}
\begin{center}
{\it The coefficients $B_r^{(m)}$ in $\langle N^s\rangle$ and an
analog of Pascal's triangle}
\end{center}
 Here we examine in some detail the coefficients appearing in the
 mean values of powers of $N$, which determine
 the intercepts $\lambda^{(2)}$ or $\lambda^{(3)}$.
  Let us specialize (\ref{34}) for $m\!=\!1,2,...,7$. Then,
e.g., for $m\!=\!5$ we have: $B_1^{(5)}\!=\!1, B_2^{(5)}\!=\!30,
B_3^{(5)}\!=\!150 , B_4^{(5)}\!=\!240, B_5^{(5)}\!=\!120$.
  Consider the correspondence between the coefficients
$B_r^{(m)}$ in $\langle N^m\rangle$ and the entries of some analog
of Pascal's triangle.
 For the classical Pascal's triangle \cite{Sloane}, there is a "rule":
each coefficient in $m$-th row equals the sum of two adjacent
coefficients in the $(m-1)$-th row.
 Now arrange our coefficients as the entries of some
triangle. Then, the triangle with entries $B_r^{(m)}$,\
$r\!=\!1,...,m$, for seven rows ($m\leq7$) looks as

\vspace{2mm}
 \hspace{30.8mm}1 \vspace{-1mm}

\hspace{25.3mm}1\hspace{11.0mm}2 \vspace{-1mm}

\hspace{20.5mm}1\hspace{9.1mm}6\hspace{11.0mm}6

\hspace{15.6mm}1 \hspace{6.4mm}14 \hspace{7.9mm}36 \hspace{7.9mm}24

\hspace{11.2mm}1 \hspace{5.4mm}30 \hspace{5.9mm}150
\hspace{6.3mm}240\hspace{6.8mm} 120

\hspace{5.9mm} 1 \hspace{3.0mm} 62 \hspace{4.2mm} 540
\hspace{5.2mm}1560 \hspace{3.8mm} 1800 \hspace{5.3mm}720

\hspace{2.4mm} 1 \hspace{1.2mm} 126 \hspace{1mm} 1806 \hspace{2.7mm}
8400 \hspace{2.2mm} 16800 \hspace{2.3mm} 15120 \hspace{3mm} 5040

\vspace{2mm}
 \noindent
 With regards to the famous Pascal's triangle, here we have two essential
distinctions. {\it First}, now we encounter unities only on the left
hand side.
 {\it Second}, the above mentioned "rule" is to be
modified. That is, each entry $B_r^{(m)}$ in $m$-th row equals the
sum of two adjacent entries above it, multiplied by $r$, the
position number of $B_r^{(m)}$:
\[
B^{(m)}_r=r\bigl(B^{(m-1)}_{r-1}+B^{(m-1)}_{r}\bigr).
\]
It is worth to note that in the mean value $\langle N^m\rangle$, the
last coefficient fixed by $r=m$ is given by the formula
\begin{equation}\vspace{-0.8mm}
B^{(m)}_{r=m} = m!\ . \label{38} %\vspace{-1mm}
\end{equation}
Here $m$ is both the power of $N$ and the row number, while $r$ is
the number of entry in $m$-th row.
 Likewise, the entries with $r=m-1$ in the $m$-th row, and
 the second entry of each $m$-th row are given as
\begin{equation} \vspace{-1mm}
B^{(m)}_{r=m-1}=\frac{(m-1)}{2} m!\ , \hspace{12mm}  B^{(m)}_{r=2} =
2 (2^{m-1}-1)=2^{m}-2.\label{39}%\vspace{-1mm}
\end{equation}
Let us note that (\ref{38}) and the first expression in (\ref{39})
coincide with the Lah numbers $L(m,1)$ and $L(m,2)$ respectively,
while the second formula in (\ref{39}) coincides with the Mersenne
numbers (see \cite{Sloane}).

\end{widetext}


\begin{thebibliography}{10}
\bibitem{footnote}                                          %34
     Recall that
     for ordinary bosons the intercept $\lambda^{(2)}$ should be constant:
     $\lambda^{(2)}_{Bose}=C^{(2)}_{Bose}-1=1$. Note also that $m$ in
     (\ref{10}) is the mass of (identical) registered particles.
\end{thebibliography}
\end{document}